\documentclass[11pt,a4paper]{JHEP3}
\usepackage{amsmath, amssymb}
\usepackage{verbatim}
\usepackage{latexsym}

\def\be{\begin{eqnarray}}
 \def\ee{\end{eqnarray}}
 \def\0{\nonumber}
\def\d{\partial}
\usepackage{euscript}

\usepackage{amsfonts}
\usepackage{verbatim}

\def\e{\epsilon}

\def\del{\partial}

\def\del{\partial}

\def\g{{\rm g}}

\def\sin{{\rm sin}}

\def\ket#1{|#1 \rangle}
\def\0{\nonumber}

\preprint{SISSA/29/2010/EP\\ \tt hep-th/yymm.xxxx}

\title{Analytic solutions for Dp branes in SFT}

\author{ L.Bonora\\
International School for Advanced Studies (SISSA)\\
Via Bonomea 265, 34136 Trieste, Italy, and INFN, Sezione di
Trieste, Italy;\\
E-mail:   \email{bonora@sissa.it},}

\author{ S.Giaccari\\
International School for Advanced Studies (SISSA)\\
Via Bonomea 265, 34136 Trieste, Italy, and INFN, Sezione di
Trieste, Italy;\\
E-mail:   \email{giaccari@sissa.it},}

\author{D.D.Tolla\\
Department of Physics and University College,
Sungkyunkwan University,
Suwon 440-746, South Korea\\
E-mail:  \email{ddtolla@skku.edu}}

\abstract{ This is the follow-up of a previous paper [ArXiv:1105.5926], where we calculated the energy of an analytic lump solution in SFT representing a D24-brane. Here we describe an analytic solution for a D$p$-brane, for any $p$, and compute its energy.}


\begin{document}

\section{Introduction}

This paper is a continuation and extension of \cite{BGT}. Recently, following an earlier suggestion
of \cite{Ellwood}, a general method has been proposed, \cite{BMT}, to
 obtain new exact analytic solutions in Witten's cubic open string field theory
(OSFT)~\cite{Witten:1985cc}, and in particular solutions that
describe inhomogeneous tachyon condensation.  There is a general
expectation that an OSFT defined on a particular boundary conformal
field theory (BCFT) has classical solutions describing other
boundary conformal field theories~\cite{Sen:1999mh,Sen:1999xm}. Previously
analytic solutions were constructed describing the tachyon
vacuum~\cite{Schnabl05, EllwoodSchnabl,Okawa1, ErlerSchnabl, RZ06, ORZ, Fuchs0,
Erler:2006hw, Erler:2006ww, Erler:2007xt, Arroyo:2010fq,
Zeze:2010jv, Zeze:2010sr, Arroyo:2010sy,Murata} and of those describing a
general marginal boundary deformations of the initial
BCFT~\cite{KORZ, Schnabl:2007az, Kiermaier:2007vu, Fuchs3,
Lee:2007ns, Kwon:2008ap, Okawa2, Okawa3, Kiermaier:2007ki,
Erler:2007rh}, see also the reviews \cite{Fuchs4, Schnabl:2010tb}. In this panorama
an element was missing: the solutions describing
inhomogeneous and relevant boundary deformations of the initial BCFT
were not known, though their existence was
predicted~\cite{Sen:1999mh,Sen:1999xm,lumps}. In \cite{Ellwood,BMT} such solutions
were put forward, in \cite{BGT,EM} the energy of a D24-brane solution was calculated for the first
time. Here we wish to extend the method and the results of \cite{BGT} to describe
analytic SFT solutions corresponding to D(25-$p$)-branes for any $p$. The extension is nontrivial because 
new aspects and problems arise for $p>1$. Apart from a greater algebraic complexity, we have 
a (new) dependence of the solutions on several (gauge) parameters and a different structure of the UV
subtractions. But the method remains essentially the same as in \cite{BGT}. The energy of the various solutions
turns out to be the expected ones.

The paper is organized as follows. In section 2 we consider a solution $\psi_{u_1,u_2}$ for a D23 brane, compute its energy functional, study its UV and IR behaviour and verify that the value of its energy functional depends on the parameter $v=u_2/u_1$. Next, in section 3, in analogy with \cite{BGT}, we introduce the $\e$-regularized solutions $\psi_{u_1,u_2}^\e$, which represents the tachyon condensation vacuum. Then we verify that
the difference $\psi_{u_1,u_2}-\psi_{u_1,u_2}^\e$, which is a solution to the equation of motion over the vacuum represented by  $\psi_{u_1,u_2}^\e$, has the expected energy of a D23-brane. At this point the extension to a generic D(25-$p$)-brane is straightforward and we summarize it in section 5.

\section{A D23-brane solution}

Let us briefly recall the technique to construct lump solutions by incorporating in SFT 
exact renormalization group flows generated in a 2D CFT by suitable relevant operators.
To start with we enlarge the well-known $K,B,c$ algebra
defined by 
\be 
K=\frac\pi2K_1^L\ket I, \quad \quad
B=\frac\pi2B_1^L\ket I,\quad\quad c= c\left(\frac12\right)\ket I,
\label{KBc} 
\ee 
in the sliver frame (obtained by mapping the UHP to
an infinite cylinder $C_2$ of circumference 2, by the sliver map
$f(z)=\frac2\pi\arctan z$), by adding a state constructed out of a (relevant) matter operator
\be 
\phi=\phi\left(\frac12\right)\ket I.\label{phi} 
\ee 
with the properties 
\be \,[c,\phi]=0,\quad\quad \,[B,\phi]=0,\quad\quad
\,[K,\phi]= \del\phi,\label{proper} 
\ee 
such that $Q$ has the following action: 
\be 
Q\phi=c\del\phi+\del c\delta\phi.
\label{actionQ} 
\ee 
One can show that 
\be \psi_{\phi}=
c\phi-\frac1{K+\phi}(\phi-\delta\phi) Bc\del c\label{psiphi} 
\ee
does indeed satisfy the OSFT equation of motion 
\be Q
\psi_{\phi}+\psi_{\phi}\psi_{\phi}=0\label{eom} 
\ee 
In order to describe the lump solution corresponding to a D24-brane in \cite{BMT,BGT} 
we used the relevant operator, \cite{Ellwood},
\be
\phi_u= u (:X^2:+2\log u+2 A)\label{phiuX}
\ee
defined on $C_1$, where $X$ is a scalar field representing the transverse space dimension, $u$ is the coupling inherited from the 2D theory and $A$ is a suitable constant.

In the case of a $D23$--brane solution, we propose, as suggested in \cite{BMT},
that the relevant operator is given by
\be
\phi_{(u_1,u_2)}= u_1 (:X_1^2:+2\log u_1+2 A)+ u_2 (:X_2^2:+2\log u_2+2 A)\label{phiu1u2}
\ee
where $X_1$ and $X_2$ are two coordinate fields corresponding to two different space directions. There is no interaction term between $X_1$ and $X_2$ in the 2D action. 

Then we require
for $\phi_u$ the following properties under the coordinate rescaling $f_t(z)=\frac zt$
\be
f_t\circ\phi_{(u_1,u_2)}(z)=\frac1t\,\phi_{(tu_1,tu_2)}\left(\frac zt\right).\label{cnd1}
\ee
The partition function corresponding to the operator (\ref{phiu1u2}) is factorized, \cite{Witten,Kutasov}:
\be 
g(u_1,u_2)=g(u_1)g(u_2)\label,\quad\quad g(u_i)= \frac 1{\sqrt {2\pi}} \sqrt{2u_i} \Gamma(2u_i)e^{2u_i(1-\log 2u_i)} 
\label{gu1u2}
\ee
where in (\ref{gu1u2}) we have already made the choice $A= \gamma -1 +\log 4\pi$.
This choice implies
\be
\lim_{u_1,u_2\to \infty} g(u_1,u_2)=1\label{g12inf}
\ee
With these properties all the non-triviality requirements of \cite{BMT,BGT} for the solution
$\psi_{(u_1,u_2)}\equiv\psi_{\phi_{(u_1,u_2)}}$ are satisfied. Therefore we can proceed to compute the energy. To this end we follow the pattern of Appendix D of \cite{BMT}, with obvious modifications.
So, for example,
\be 
\langle X_1^2(\theta)  X_2^2(\theta')\rangle_{Disk} = \langle X_1^2(\theta)\rangle_{Disk} \langle X_2^2(\theta')\rangle_{Disk}= Z(u_1)h_{u_1} Z(u_2)h_{u_2} 
\ee and so on.

Going through the usual derivation one gets that the energy functional is given by  
\be
E[\psi_{(u_1,u_2)}]= -\frac 16 \langle \psi_{(u_1,u_2)} \psi_{(u_1,u_2)}\psi_{(u_1,u_2)}\rangle\label{E121}
\ee
where
\be
 E[\psi_{(u_1,u_2)}] &=&\frac 16\int_{0}^\infty
dt_1dt_2dt_3\,{\cal E}_0(t_1,t_2,t_3)\,g(u_1T,u_2T)\0\\
&&\cdot\Bigg\{8u_1^3\,G_{2u_1T}(\frac{2\pi t_1}T)G_{2u_1T}(\frac{2\pi
(t_1+t_2)}T)G_{2u_1T}(\frac{2\pi t_2}T)\0\\
&&+ 8u_2^3\,G_{2u_2T}(\frac{2\pi t_1}T)G_{2u_2T}(\frac{2\pi
(t_1+t_2)}T)G_{2u_2T}(\frac{2\pi t_2}T)\0\\
&&+\Biggl(2u_1^3 \Big(-\frac{\partial_{u_1T}g(u_1T,u_2T)}{g(u_1T,u_2T)}\Big)
+2u_1^2 u_2\Big(-\frac{\partial_{u_2T}g(u_1T,u_2T)}{g(u_1T,u_2T)}\Big)\Biggr)\0\\
&&~~~~\cdot \Big(G_{2u_1T}^2(\frac{2\pi
t_1}T)+G_{2u_1T}^2(\frac{2\pi (t_1+t_2)}T)+G_{2u_1T}^2(\frac{2\pi t_2}T)\Big)\0\\
&&+\Biggl(2u_2^3 \Big(-\frac{\partial_{u_2T}g(u_1T,u_2T)}{g(u_1T,u_2T)}\Big)
+2u_2^2 u_1\Big(-\frac{\partial_{u_1T}g(u_1T,u_2T)}{g(u_1T,u_2T)}\Big)\Biggr)\0\\
&&~~~~\cdot \Big(G_{2u_2T}^2(\frac{2\pi
t_1}T)+G_{2u_2T}^2(\frac{2\pi (t_1+t_2)}T)+G_{2u_2T}^2(\frac{2\pi t_2}T)\Big)\0\\
&&+ \Biggl(u_1 \Big(-\frac{\partial_{u_1T}g(u_1T,u_2T)}{g(u_1T,u_2T)}\Big)+
u_2\Big(-\frac{\partial_{u_2T}g(u_1T,u_2T)}{g(u_1T,u_2T)}\Big) \Biggr)^3\Bigg\}\label{E122}
\ee
When writing $ \partial_{u_1T}g(u_1T,u_2T) $ we mean  that we differentiate (only) with respect to the first entry, and when $ \partial_{u_2T}g(u_1T,u_2T) $ (only) with respect to the second. This can be written also as
\be
  E[\psi_{(u_1,u_2)}]&=&\frac 16\int_{0}^\infty
dt_1dt_2dt_3\,{\cal E}_0(t_1,t_2,t_3)\,g(u_1T,u_2T)\0\\
&&\cdot\Bigg\{8u_1^3\,G_{2u_1T}(\frac{2\pi t_1}T)G_{2u_1T}(\frac{2\pi
(t_1+t_2)}T)G_{2u_1T}(\frac{2\pi t_2}T)\0\\
&&+ 8u_2^3\,G_{2u_2T}(\frac{2\pi t_1}T)G_{2u_2T}(\frac{2\pi
(t_1+t_2)}T)G_{2u_2T}(\frac{2\pi t_2}T)\0\\
&&+\Bigl(-2u_1^2 \frac{\partial_{T}g(u_1T,u_2T)}{g(u_1T,u_2T)} \Bigr)\0\\
&&~~~~\cdot \Big(G_{2u_1T}^2(\frac{2\pi
t_1}T)+G_{2u_1T}^2(\frac{2\pi (t_1+t_2)}T)+G_{2u_1T}^2(\frac{2\pi t_2}T)\Big)\0\\
&&+\Bigl(-2u_2^2  \frac{\partial_{T}g(u_1T,u_2T)}{g(u_1T,u_2T)}\Bigr) \0\\
&&~~~~\cdot \Big(G_{2u_2T}^2(\frac{2\pi
t_1}T)+G_{2u_2T}^2(\frac{2\pi (t_1+t_2)}T)+G_{2u_2T}^2(\frac{2\pi t_2}T)\Big)\0\\
&&+ \Bigl( -\frac{\partial_{T}g(u_1T,u_2T)}{g(u_1T,u_2T)} \Bigr)^3\Bigg\}
\ee
where now $\partial_{T}g(u_1T,u_2T)$ means differentiation with respect to both entries. A further useful form is  the following one
\be
E[\psi_{(u_1,u_2)}]&=&\frac16 \int_0^\infty ds\; s^2\int_0^1
dy\int_0^{y} dx\,\frac 4\pi \,\sin\pi x\,\sin\pi y\,\sin\pi(x-y)\g(s,v s)\label{Efinal}\\
&&\cdot\Bigg\{\,G_{s}(2\pi x)G_{s}(2\pi (x-y))G_{s}(2\pi y)\0\\
&&+ v^3\,G_{v s}(2\pi x)G_{vs}(2\pi (x-y))G_{vs}(2\pi y)\0\\
&&-\frac 12\Bigl( \frac{\partial_{s}\g(s,vs)}{\g(s,vs)} \Bigr)\Big(G_{s}^2(2\pi x)+G_{s}^2(2\pi (x-y))+G_{s}^2(2\pi y)\Big)\0\\
&&-\frac 12\Bigl( v^2  \frac{\partial_{s}\g(s,vs)}{\g(s,vs)}\Bigr)\Big(G_{vs}^2(2\pi x)+G_{vs}^2(2\pi (x-y))+G_{vs}^2(2\pi y)\Big)\0\\
&&+ \Bigl( -\frac{\partial_{s}\g(s,vs)}{\g(s,vs)} \Bigr)^3\Bigg\}\label{E123}
\ee
where $s=2u_1T, v=\frac {u_2}{u_1}$ and, by definition, $\g(s,vs)\equiv g(s/2,vs/2)= g(u_1T,u_2T)$. The derivative $\partial_s$ in $\partial_{s}\g(s,vs)$ acts on both entries.
We see that, contrary to \cite{BMT}, where the $u$ dependence was completely absorbed within
the integration variable, in (\ref{E123}) there is an explicit dependence on $v$.  

\subsection{The IR and UV behaviour}

First of all we have to find out whether $E[\psi_{(u_1,u_2)}]$ is finite and whether it depends on $v$.

To start with let us notice that the structure of the $x,y$ dependence is the same
as in \cite{BGT}. Therefore we can use the results already found there, with exactly the 
same IR ($s\to \infty$) and UV ($s\approx 0$) behaviour. The differences with \cite{BGT} come from the various factors containing $\g$ or derivatives thereof. The relevant IR asymptotic behaviour is
\be
g(s,vs)\approx    1+\frac {1+v}{24 v }\frac 1s \label{gasympt}
\ee
for large $s$ ($v$ is kept fixed to some positive value).
The asymptotic behaviour does not change with respect to the D24-brane case (except perhaps
for the overall dominant asymptotic coefficient, which is immaterial as far as integrability is concerned), so we can conclude that the integral in (\ref{E123}) is convergent for large 
$s$, where the overall integrand behaves asymptotically as $ 1/s^2$.  
 
Let us come next to the UV behaviour ($s\approx 0$).
To start with let us consider the term not containing $G_s$. We have
\be
&&\frac 1{4\pi^2}s^2 g(s,vs) \left(\frac {\d_s g(s,vs)}{g(s,vs)}\right)^3=
-\frac{1}{16 \left(\pi ^3 \sqrt{v}\right) s^2}\0\\
&&-\frac{1}{8 \pi ^3 \sqrt{v} s}\bigl((1+v)(1+2 \gamma)+2 \log 2+2(1+v) \log s+2 v \log (2 v)\bigr) +{\cal O}((\log s)^2)\label{ord0IR}
\ee
The double pole in zero is to be expected. Once we integrate over $s$ we obtain a behaviour 
$\sim \frac 1s$ near $s=0$. This singularity corresponds to $\sim\delta(0)^2\sim V^2$, which can be interpreted as the D25 brane energy density multiplied by the square of the (one-dimensional) volume, see Appendix C of \cite{BGT}). In order to extract a finite quantity from the integral (\ref{E123}) we have to subtract this singularity. We proceed as 
in  \cite{BGT} and find that the function to be subtracted to the LHS of (\ref{ord0IR}) is
\be
h_1(v,s)&=& \Big(-\frac{1}{16 \left(\pi ^3 \sqrt{v}\right) s^2}+\frac{1}{16 \pi ^3 \sqrt{v} s}\0\\
&&-\frac{1}{8 \pi ^3 \sqrt{v} s}\bigl((1+v)(1+2 \gamma)+2 \log 2+2(1+v) \log s+2 v \log (2 v) \bigr)\Big)\0\\
&&\cdot\frac{e^{\frac{s}{s^2-1}} \left(1+2 s-2 s^2+2s^3+s^4\right)}{\left(-1+s^2\right)^2}
\label{subtrh1}
\ee
in the interval $0\leq s\leq 1$ and 0 elsewhere. It is important to remark that both the singularity and the subtraction are $v$-dependent.

As for the quadratic terms in $G_s$ and $G_{vs}$ the overall UV singularity is
\be
-\frac{3}{16 \left(\pi ^3 \sqrt{v}\right) s^2}-\frac{3 (1+v)}{8 \left(\pi ^3 \sqrt{v}\right) s}+{\cal O}((\log s)^2)\label{ord2IR}
\ee
and the corresponding function to be subtracted from the overall integrand is
\be
h_2(v,s)= -\frac{3 e^{\frac{s}{s^2-1}} \left(1+2 s-2 s^2+2 s^3+s^4\right) (1+s+2 s v)}{16 \pi ^3 s^2 \left(s^2-1\right)^2 \sqrt{v}}\label{subtrh2}
\ee
in the interval $0\leq s\leq 1$ and 0 elsewhere. Also in this case the subtraction is $v$ dependent.

Finally let us come to the cubic term in $G_s$ and $G_{vs}$. Altogether the UV singularity due to the cubic terms is
\be
-\frac{1}{8 \left(\pi ^3 \sqrt{v}\right) s^2}+\frac{(\gamma+\log s)(1+v)+\log 2+v \log(2v)}{4 \pi ^3 \sqrt{v} s}+ {\cal O}((\log s)^2)\label{ord3IR}
\ee
The overall function we have to subtract from the corresponding integrand is
\be
h_3(v,s)&=& \frac{2}{16\pi ^3 \sqrt{v} s}\frac{e^{\frac{s}{s^2-1}} \left(1+2 s-2 s^2+2s^3+s^4\right)}{\left(s^2-1\right)^2}\0\\
&&\cdot \bigl(-1+ s+2s (1+v)(\gamma+ \log s) + s \log 4 +2 s v \log(2 v)\bigr)\label{subtrh3}
\ee
for $0\leq s\leq 1$ and 0 elsewhere. Also in this case the subtraction is $v$ dependent.

As explained in \cite{BGT} the result of all these subtractions does not depend on the particular functions
$h_1,h_2,h_3$ we have used, provided the latter satisfy a few very general criteria.

After all these subtractions the integral in (\ref{E123}) is finite, but presumably
$v$ dependent. This is confirmed by a numerical analysis. For instance, for
$v=1$ and 2 we get $E^{(s)}[\psi_{(u_1,u_2)}]=0.0892632$ and  0.126457, respectively,
where the superscript $^{(s)}$ means UV subtracted. It is clear that this cannot represent a physical energy. This is not surprising. We have already remarked in \cite{BGT} that the UV subtraction procedure carries with itself a certain amount of arbitrariness. Here we have in addition an explicit $v$ dependence that renders this fact even more clear. The way out is the same as in \cite{BGT}. We will compare the (subtracted) energy of $\psi_{(u_1,u_2)}$
with the (subtracted) energy of a solution representing the tachyon condensation vacuum,
and show that the result is independent of the subtraction scheme.

\section{The $\epsilon$-regularization} 
 
As we did in section 8 of \cite{BGT}, we need to introduce the $\epsilon$-regularization and the  $\e$-regularized solution corresponding to
(\ref{phiu1u2}). We recall the general form of such solution
 \be
\psi_{\phi}= c(\phi+\epsilon) - \frac 1{K+\phi+\epsilon} (\phi+\epsilon -\delta \phi) Bc\partial c
\label{psiphieps}
\ee
where $\e$ is an arbitrary small number. In the present case  
\be
\phi\equiv \phi_{(u_1,u_2)}= u_1 (:X_1^2:+2\log u_1+2 A)+ u_2 (:X_2^2:+2\log u_2+2 A)\label{phiphiu1u2}
\ee
It is convenient to split $\e=\e_1+\e_2$ and associate $\e_1$ to the first piece in the RHS of (\ref{phiphiu1u2}) and $\e_2$ to the second. We will call the corresponding solution
$\psi_{(u_1,u_2)}^\e$. After the usual manipulations the result is
\be
E[\psi_{(u_1,u_2)}^\e] &=&\frac 16\int_{0}^\infty
dt_1dt_2dt_3\,{\cal E}_0(t_1,t_2,t_3)\,g(u_1T,u_2T)\, e^{-\e T}\0\\
&&\cdot\Bigg\{8u_1^3\,G_{2u_1T}(\frac{2\pi t_1}T)G_{2u_1T}(\frac{2\pi
(t_1+t_2)}T)G_{2u_1T}(\frac{2\pi t_2}T)\0\\
&&+ 8u_2^3\,G_{2u_2T}(\frac{2\pi t_1}T)G_{2u_2T}(\frac{2\pi
(t_1+t_2)}T)G_{2u_2T}(\frac{2\pi t_2}T)\0\\
&&+\Biggl(2u_1^3 \Big(\frac{\e_1}{u_1}-\frac{\partial_{u_1T}g(u_1T,u_2T)}{g(u_1T,u_2T)}\Big)
+2u_1^2 u_2\Big(\frac{\e_2}{u_2}-\frac{\partial_{u_2T}g(u_1T,u_2T)}{g(u_1T,u_2T)}\Big)\Biggr)\0\\
&&~~~~\cdot \Big(G_{2u_1T}^2(\frac{2\pi
t_1}T)+G_{2u_1T}^2(\frac{2\pi (t_1+t_2)}T)+G_{2u_1T}^2(\frac{2\pi t_2}T)\Big)\0\\
&&+\Biggl(2u_2^3 \Big(\frac{\e_2}{u_2}-\frac{\partial_{u_2T}g(u_1T,u_2T)}{g(u_1T,u_2T)}\Big)
+2u_2^2 u_1\Big(\frac{\e_1}{u_1}-\frac{\partial_{u_1T}g(u_1T,u_2T)}{g(u_1T,u_2T)}\Big)\Biggr)\0\\
&&~~~~\cdot \Big(G_{2u_2T}^2(\frac{2\pi
t_1}T)+G_{2u_2T}^2(\frac{2\pi (t_1+t_2)}T)+G_{2u_2T}^2(\frac{2\pi t_2}T)\Big)\0\\
&&+ \Biggl(u_1 \Big(\frac{\e_1}{u_1}-\frac{\partial_{u_1T}g(u_1T,u_2T)}{g(u_1T,u_2T)}\Big)+
u_2\Big(\frac{\e_2}{u_2}-\frac{\partial_{u_2T}g(u_1T,u_2T)}{g(u_1T,u_2T)}\Big) \Biggr)^3\Bigg\}\label{reg1}
\ee
or
\be
E[\psi_{(u_1,u_2)}^\e] &=&\frac 16\int_{0}^\infty
dt_1dt_2dt_3\,{\cal E}_0(t_1,t_2,t_3)\,g(u_1T,u_2T)\, e^{-\e T}\0\\
&&\cdot\Bigg\{8u_1^3\,G_{2u_1T}(\frac{2\pi t_1}T)G_{2u_1T}(\frac{2\pi
(t_1+t_2)}T)G_{2u_1T}(\frac{2\pi t_2}T)\0\\
&&+ 8u_2^3\,G_{2u_2T}(\frac{2\pi t_1}T)G_{2u_2T}(\frac{2\pi
(t_1+t_2)}T)G_{2u_2T}(\frac{2\pi t_2}T)\0\\
&&+\Biggl(2u_1^2\Big(\e- \frac{\partial_{T}g(u_1T,u_2T)}{g(u_1T,u_2T)} \Big)\Biggr)\0\\
&&~~~~\cdot \Big(G_{2u_1T}^2(\frac{2\pi
t_1}T)+G_{2u_1T}^2(\frac{2\pi (t_1+t_2)}T)+G_{2u_1T}^2(\frac{2\pi t_2}T)\Big)\0\\
&&+\Biggl(2u_2^2\Big(\e-  \frac{\partial_{T}g(u_1T,u_2T)}{g(u_1T,u_2T)}\Big)\Bigr) \0\\
&&~~~~\cdot \Big(G_{2u_2T}^2(\frac{2\pi
t_1}T)+G_{2u_2T}^2(\frac{2\pi (t_1+t_2)}T)+G_{2u_2T}^2(\frac{2\pi t_2}T)\Big)\0\\
&&+ \Bigl(\e -\frac{\partial_{T}g(u_1T,u_2T)}{g(u_1T,u_2T)} \Bigr)^3\Bigg\}\label{reg2}
\ee
and finally
\be
E[\psi^\e_{(u_1,u_2)}]&=&\frac16 \lim_{\e \to 0} \int_0^\infty ds\; s^2\int_0^1
dy\int_0^{y} dx\,{\cal E}(1-y,x)\, \g(s,v s)\, e^{-\eta s}\label{Efinalepsilon}\\
&&\cdot\Bigg\{\,G_{s}(2\pi x)G_{s}(2\pi (x-y))G_{s}(2\pi y)\0\\
&&+ v^3\,G_{v s}(2\pi x)G_{vs}(2\pi (x-y))G_{vs}(2\pi y)\0\\
&&+\frac 12\Bigl( \eta-\frac{\partial_{s}\g(s,vs)}{\g(s,vs)} \Bigr)\Big(G_{s}^2(2\pi x)+G_{s}^2(2\pi (x-y))+G_{s}^2(2\pi y)\Big)\0\\
&&+\frac 12 v^2\Bigl(\eta-\frac{\partial_{s}\g(s,vs)}{\g(s,vs)}\Bigr)\Big(G_{vs}^2(2\pi x)+G_{vs}^2(2\pi (x-y))+G_{vs}^2(2\pi y)\Big)\0\\
&&+ \Bigl(\eta-\frac{\partial_{s}\g(s,vs)}{\g(s,vs)} \Bigr)^3\Bigg\}\0
\ee
where ${\cal E}(1-y,x)=\frac4\pi \,\sin\pi x\,\sin\pi y\,\sin\pi(x-y)\0$ and $\eta=\frac{\e}{2u_1}$. It is worth remarking that the result (\ref{Efinalepsilon}) does not depend on the splitting $\e=\e_1+\e_2$.

The integrand in (\ref{Efinalepsilon}) has the same leading singularity in the UV as the integrand
of (\ref{E123}). The subleading singularity on the other hand may depend on $\e$.
Thus it must undergo an UV subtraction that generically depends on $\e$. We will denote the corresponding subtracted integral by $E^{(s)}[\psi^\e_{(u_1,u_2)}]$. The important remark here is, however, that
in the limit $\e\to 0$ both (\ref{Efinalepsilon}) and (\ref{E123}) undergo the same subtraction.

The factor of $ e^{-\eta s}$ appearing in the integrand of (\ref{Efinalepsilon}) changes completely its IR structure. It is in fact responsible for cutting out the contribution at infinity that characterizes (\ref{E123}) and (modulo the arbitrariness in the UV subtraction) makes up the energy of the D23 brane. 

In keeping with \cite{BGT}, we interpret $\psi^\e_{(u_1,u_2)}$ as a tachyon condensation vacuum solution
and $E^{(s)}[\psi^\e_{(u_1,u_2)}]$ the energy of such vacuum. This energy is actually $v$- (and possibly $\e$)-dependent. We will explain later on how it can be set to 0.

\section{The energy of the $D23$--brane}
\vskip 0.5cm

As explained in \cite{BGT}, the problem of finding the right energy of the D23 brane consists in constructing a solution over the vacuum represented by
$\psi_{(u_1,u_2)}^\e$ (the tachyon condensation vacuum).
The equation of motion at such vacuum is
\begin{align}\label{EOMTV}
{\cal Q}\Phi+\Phi\Phi=0,\quad {\rm where}~~{\cal
Q}\Phi=Q\Phi+\psi_{(u_1,u_2)}^\e\Phi+\Phi \psi_{(u_1,u_2)}^\e
\end{align}
One can easily show that
\begin{align}\label{psiupsie}
\Phi_0= \psi_{(u_1,u_2)}-\psi_{(u_1,u_2)}^\e
\end{align}
is a solution to (\ref{EOMTV}). The action at the tachyon vacuum is
\begin{align}\label{TVaction}
-\frac12\langle{\cal
Q}\Phi_0,\Phi_0\rangle-\frac13\langle\Phi_0,\Phi_0\Phi_0\rangle.
\end{align}
Thus the energy is
\be
E[\Phi_0]&=&-\frac16\langle\Phi_0,\Phi_0\Phi_0\rangle
=-\frac16\big[\langle\psi_{(u_1,u_2)} ,\psi_{(u_1,u_2)}\psi_{(u_1,u_2)}\rangle
-\langle\psi_{(u_1,u_2)}^\e, \psi_{(u_1,u_2)}^\e\psi_{(u_1,u_2)}^\e\rangle\0\\
&&-3\langle \psi_{(u_1,u_2)}^\e, \psi_{(u_1,u_2)}\psi_{(u_1,u_2)}\rangle+3\langle\psi_{(u_1,u_2)},\psi_{(u_1,u_2)}^\e\psi_{(u_1,u_2)}^\e\rangle\big].\label{EPhi0}
\ee
Eq.(\ref{psiupsie}) is the lump solution at the tachyon vacuum, therefore this energy must be the energy of the lump.

The two additional terms $\langle \psi_{(u_1,u_2)}^\e,\psi_{(u_1,u_2)}\psi_{(u_1,u_2)}\rangle$
and $\langle\psi_{(u_1,u_2)},\psi_{(u_1,u_2)}^\e\psi_{(u_1,u_2)}^\e\rangle$ are given by
\be
&& \langle \psi_{(u_1,u_2)}^\e,\psi_{(u_1,u_2)}\psi_{(u_1,u_2)}\rangle=- \lim_{\e\to 0}\int_0^\infty ds\;
s^2\int_0^1 dy\int_0^{y} dx\,e^{-\eta s}{\cal
E}(1-y,x) \,e^{\eta sy}\, \g(s,vs)\0\\
&&\quad\quad\cdot\Bigg\{\Big(\eta -\frac{\partial_{s}\g(s,vs)}{\g(s,vs)}\Big)\Big(-\frac{\partial_{s}\g(s,vs)}{\g(s,vs)}\Big)^2\label{euu}\\
&&\quad\quad +G_{s}(2\pi x)G_{s}(2\pi (x-y))G_{s}(2\pi y) + v^3\,G_{v s}(2\pi x)G_{vs}(2\pi (x-y))G_{vs}(2\pi y)\0\\
&&\quad\quad+\frac 12\Bigl( \eta-\frac{\partial_{s}\g(s,vs)}{\g(s,vs)} \Bigr)\Bigl(G_{s}^2(2\pi(x))+v^2 G_{vs}^2(2\pi(x))\Bigr)\0\\
&&\quad\quad+ \frac 12\Big(-\frac{\partial_{s}\g(s,vs)}{\g(s,vs)}\Big)\Bigl(G_{s}^2(2\pi
y)+G_{s}^2(2\pi (x-y))+ v^2 \Big(G_{vs}^2(2\pi
y)+G_{vs}^2(2\pi (x-y) )\Big)\Bigr) \Bigg\}.\0
\ee
and
\be
&&\langle\psi_{(u_1,u_2)},\psi_{(u_1,u_2)}^\e\psi_{(u_1,u_2)}^\e\rangle=- \lim_{\e\to 0}\int_0^\infty ds\;
s^2\int_0^1 dy\int_0^{y} dx\,e^{-\eta s}{\cal
E}(1-y,x) \,e^{\eta sx}\, \g(s,vs) \0\\
&&\quad\quad\cdot\Bigg\{\Big(\eta -\frac{\partial_{s}\g(s,vs)}{\g(s,vs)}\Big)^2\Big(-\frac{\partial_{s}\g(s,vs)}{\g(s,vs)}\Big)\label{uee}\\
&&\quad\quad +G_{s}(2\pi x)G_{s}(2\pi (x-y))G_{s}(2\pi y) + v^3\,G_{v s}(2\pi x)G_{vs}(2\pi (x-y))G_{vs}(2\pi y)\0\\
&&\quad\quad+\frac 12\Bigl( \eta-\frac{\partial_{s}\g(s,vs)}{\g(s,vs)}\Bigr) \Bigl(G_{s}^2(2\pi
x)+G_{s}^2(2\pi y)+ v^2 \Big(G_{vs}^2(2\pi
x)+G_{vs}^2(2\pi y)\Big)\Bigr)\0\\
&&\quad\quad+ \frac 12\Big(-\frac{\partial_{s}\g(s,vs)}{\g(s,vs)}\Big)\Bigl(G_{s}^2(2\pi(x-y))+v^2 G_{vs}^2(2\pi(x-y))\Bigr) \Bigg\}.\0
\ee
Now we insert in (\ref{EPhi0}) the quantities we have just computed together with
(\ref{E123}) and (\ref{Efinalepsilon}). We have of course to subtract their UV singularities.
As we have already remarked above, such subtractions are the same for all terms in (\ref{EPhi0}) in the limit $\e\to 0$, therefore they cancel out. So the result we obtain from
(\ref{EPhi0}) is subtraction-independent and we expect it to be the physical result.

In fact the expression we obtain after the insertion of (\ref{E123},\ref{Efinalepsilon},\ref{euu}) and (\ref{uee}) in (\ref{EPhi0}) looks very complicated. But it simplifies drastically in the limit $\e\to 0$.  
As was noticed in \cite{BGT}, in this limit we can drop the factors $e^{\eta sx}$ and $e^{\eta sy}$ in (\ref{euu}) and (\ref{uee}) because of continuity\footnote{It is useful to recall that the limit $\e\to 0$ can be taken safely inside the integration only if the integral without the factor $e^{\eta sx}$ or $e^{\eta sy}$ is convergent. This is true for the $x$ and $y$ integration, but it is not the case for instance for the integral (\ref{1st}) below.}. What we cannot drop {\it a priori}
is the factor $e^{-\eta s}$.

Next it is convenient to introduce $\tilde \g(s,vs)= e^{-\eta s}\g(s,vs)$ and notice that 
\be
\eta-\frac{\partial_{s}\g(s,vs)}{\g(s,vs)}= - \frac{\partial_{s}\tilde\g(s,vs)}{\tilde\g(s,vs)}\label{useful1}
\ee
Another useful simplification comes from the fact that (without the $e^{\eta sx}$ or $e^{\eta sy}$ factors) upon integrating over $x,y$ the three terms proportional to $G_{s}^2(2\pi x)$, $G_{s}^2(2\pi y)$ and $G_{s}^2(2\pi (x-y))$, respectively,  give rise to the same contribution. With this in mind one can easily realize that most of the terms cancel and what remains is
\be
E[\Phi_0]&=& \frac16 \lim_{\e \to 0} \int_0^\infty ds\; s^2\int_0^1
dy\int_0^{y} dx\,{\cal E}(1-y,x)\,\Bigg\{ \g(s,v s)\, (1-e^{-\eta s})\0\\
&&\cdot\Bigg[G_{s}(2\pi x)G_{s}(2\pi (x-y))G_{s}(2\pi y)
+ v^3\,G_{v s}(2\pi x)G_{vs}(2\pi (x-y))G_{vs}(2\pi y)\0\\
&&+\frac 12\Bigl(-\frac{\partial_{s}\g(s,vs)}{\g(s,vs)} \Bigr)\Big(G_{s}^2(2\pi x)+G_{s}^2(2\pi (x-y))+G_{s}^2(2\pi y)\Big)\0\\
&&+\frac 12 v^2\Bigl(\eta-\frac{\partial_{s}\g(s,vs)}{\g(s,vs)}\Bigr)\Big(G_{vs}^2(2\pi x)+G_{vs}^2(2\pi (x-y))+G_{vs}^2(2\pi y)\Big)\0\\
&&+ \Bigl(-\frac{\partial_{s}\g(s,vs)}{\g(s,vs)} \Bigr)^3\Bigg]\0\\
&&+\tilde \g(s,v s) \Bigl(\frac{\partial_{s}\tilde\g(s,vs)}{\tilde\g(s,vs)} -\frac{\partial_{s}\g(s,vs)}{\g(s,vs)} \Bigr)^3\Bigg\}\label{Eultimate}
\ee
The term proportional to $1-e^{-\eta s}$ vanishes in the limit $\e\to 0$ because the integral
without this factor is finite (after UV subtraction). Therefore we are left with
\be
E[\Phi_0]&=& \frac16 \lim_{\e \to 0} \int_0^\infty ds\; s^2\int_0^1
dy\int_0^{y} dx\,{\cal E}(1-y,x)\, \g(s,v s)\, e^{-\eta s}\, \eta^3 \0\\
&=&\frac 1{4\pi^2} \lim_{\e\to 0} \int_0^{\infty} ds\,s^2 \g(s,vs) \eta^3 e^{-\eta s}.\label{1st}
\ee
where $\g(s,vs)=\g(s)\g(vs)$.
In \cite{BGT} the analog of this was the coefficient $\alpha$ that determines the energy of the solution. 
This contribution comes from the $\e^3$ term in the last line of  (\ref{reg2}).
If one knows the asymptotic expansion of the integrand for large $s$, it is very easy to extract the exact $\e\to 0$ result of the integral. We recall that the UV singularity
has been subtracted away, therefore the only nonvanishing contribution to the integral (\ref{1st}) may come from $s\to \infty$. In fact splitting the $s$ integration as $0\leq s\leq M$ and $M\leq s <\infty$, where $M$ is a very large number, it is easy to see the the integration in the first interval vanishes in the limit $\e\to 0$. As for the second integral we have to use the asymptotic expansion of $\g(s,vs)$: $\g(s,vs)\approx    1+\frac {1+v}{12 v }\frac 1s +\ldots $.  Integrating term by term from $M$ to $\infty$, the dominant one gives
\be
 \frac 1{4\pi^2}e^{-\eta M } (2 +2 M \eta+M^2 \eta^2) \label{0term1}
\ee
which, in the  $\e\to 0$ limit, yields $\frac 1{2\pi^2}$. The other terms are irrelevant in the $\e\to 0$ limit.
Therefore we have
\be\label{EPhi0result}
 E[\Phi_0]=\frac 1{2\pi^2}.
\ee
We recall that 1 in the numerator on the RHS is to be identified with $\lim_{s\to\infty} g(s,vs)$.

We conclude that 
\be 
T_{23}= \frac 1{2\pi^2}\label{T23}
\ee
This is the same as $T_{24}$, so it may at first be surprising. But in fact it is correct because of the normalization discussed in App. C of \cite{BGT}. Compare with eqs.(C.1) and (C.7) there: when we move from a $Dp$-brane to a $D(p-1)$-brane, the tension is multiplied by $2\pi $ (remember that $\alpha'=1$), but simultaneously we have to divide by $2\pi$ because the volume is measured with units differing by $2\pi$ (see after eq. (C.6)).

In more detail the argument goes as follows (using the notation of Appendix C of \cite{BGT}). The volume in our normalization is $V=2\pi {\EuScript V}$, where ${\EuScript V}$ is the volume in Polchinski's textbook normalization, \cite{Polchinski}, see also \cite{Okawa02}. The energy functional
for the D24 brane is proportional to the 2D zero mode normalization (which determines the normalization of the partition function). The latter is proportional to $\frac 1V$. 
Since $V=2\pi {\EuScript V}$, normalizing with respect to ${\EuScript V}$ is equivalent to multiplying the energy by $2\pi$. This implies that
\be 
T_{D24}= \frac 1{2\pi} {\EuScript T}_{D24}\label{T24}
\ee
where ${\EuScript T}$ represents the tension in Polchinski's units.
The energy functional in (\ref{Efinal}) depends linearly on the normalization of $\g(s,vs)$, which is the square of the normalization of $g(s)$, so is proportional to $\frac 1 {V^2}$. Therefore the ratio between the energy
with the two different zero mode normalizations is $(2\pi)^2$. Consequently we have
\be 
T_{D23}= \frac 1{(2\pi)^2} {\EuScript T}_{D23}\label{Ttau23}
\ee
Since, from Polchinski, we have
\be 
{\EuScript T}_{D23}= 2\pi\, {\EuScript T}_{D24}=  (2\pi)^2\, {\EuScript T}_{D25}= 2\label{D23D24}
\ee
eq.(\ref{T23}) follows. 

We end this section with two comments. The first is about $E^{(s)}[\psi^\e_{u_1,u_2}]$. This is interpreted as the energy
at the tachyon condensation vacuum after the UV subtraction, which represents itself the energy of the tachyon condensation vacuum. Therefore it should vanish. In fact it does not vanish and its value is $(v,\e)$-dependent. The reason it does not vanish is that the subtraction itself is $(v,\e)$-dependent and this is due to the arbitrariness of the subtraction scheme. However we can always fix $E^{(s)}[\psi^\e_{u_1,u_2}]$ to zero by subtracting a 
suitable constant. Of course we have to subtract the same constant from $E^{(s)}[\psi_{u_1,u_2}]$.

The second comment concerns the dependence on $\e$ of (\ref{EPhi0}). The result we have derived in this section
makes essential use of the limit $\e\to 0$, but we believe that it should hold for any $\e$. In \cite{BGT}
it was in fact argued that this should be so, based on the $\e$--independence of the UV subtractions. In this paper the UV subtractions are generically $\e$-dependent and we cannot use the same simple argument. However
there is no reason to believe that the RHS of (\ref{EPhi0}) is $\e$ dependent, although it is more complicated to prove it. Such complication has to do only with the technicalities of the $\e$-regularization. It is possible to envisage other regularizations in which the UV subtractions are independent of the regulator. We will pursue this point elsewhere.

\section{D(25-p) brane solutions}

The previous argument about D-brane tensions can be easily continued and we always find that the value to be expected is
\be
T_{25-p}= \frac 1{2\pi^2}, \quad\quad \forall p\geq 1\label{T25p}
\ee
An analytic solution with such energy is easily found. We introduce the relevant operator
\be
\phi_u= \sum_{i=1}^p u_i (:X_i^2:+2\log u_i+2 A)\label{phiuiX}
\ee
where $X_i$ will represent the transverse direction to the brane and $u_i$ the corresponding 2D couplings.  
Since the $u_i$ couplings evolve independently and linearly, the partition function 
will be  $g(u_1,\ldots, u_p)= g(u_1)g(u_2)\ldots g(u_p)$. 

The derivation of the energy of such solutions is a straightforward generalization of the one above for the D23-brane and we will not repeat it.
The final result for the energy above the tachyon condensation vacuum is
\be
E[\Phi_0]&=& \frac16 \lim_{\e \to 0} \int_0^\infty ds\; s^2\int_0^1
dy\int_0^{y} dx\,{\cal E}(1-y,x)\, \g(s,v_1 s,\ldots,v_{p-1}s)\, e^{-\eta s}\, \eta^3 \0\\
&=&-\frac 1{4\pi^2} \lim_{\e\to 0} \int_0^{\infty} ds\,s^2 \g(s,v_1s,\ldots,v_{p-1}s) \eta^3 e^{-\eta s}.\label{2nd}
\ee
where $v_1=\frac {u_2}{u_1}, v_2=\frac {u_3}{u_1},\ldots$. It is understood that the UV singularity has 
been subtracted away from the integral in the RHS, therefore the only contribution comes from the region
of large $s$. 
Since, again
$\lim_{s\to \infty} \g( s,v_1s,\ldots,v_{p-1}s)=1$, we find straightaway that
\be\label{alphasecond}
E[\Phi_0]= \frac 1{2\pi^2}.
\ee
from which (\ref{T25p}) follows.

\end{document}